\documentclass[]{IEEEtran}
\usepackage{cite}
\usepackage{amsmath,amssymb,amsfonts}
\usepackage{algorithmic}
\usepackage{graphicx}
\usepackage{textcomp}
\usepackage{xcolor}

\ifCLASSOPTIONcompsoc
\usepackage[caption=false, font=normalsize, labelfont=sf, textfont=sf]{subfig}
\else
\usepackage[caption=false, font=footnotesize]{subfig}
\fi

\usepackage{url}								
\usepackage{siunitx}							
\usepackage{booktabs}							
\usepackage{tabularx}
\newcolumntype{Y}{>{\centering\arraybackslash}X} 

\usepackage{makecell}							
\usepackage{multirow}							
\usepackage[display-foreign=false]{acro}			
\usepackage{gensymb}
\acsetup{single}								

\usepackage{flushend} 				
\usepackage{tikz}								

\usepackage{soul}
\usepackage[most]{tcolorbox}

\DeclareAcronym{LCMP}{
	short = LCMP,
	long = linear constrained minimum power 
}

\begin{document}
	
\title{Multi-Objective Distributed Beamforming Using High-Accuracy Synchronization and Localization}

\author{ Ahona Bhattacharyya,~\IEEEmembership{Graduate~Student~Member,~IEEE,} and Jeffrey A. Nanzer,~\IEEEmembership{Senior Member,~IEEE}%
	\thanks{This work was supported in part by the Defense Advanced Research Projects Agency under grant \#HR00112010015, the Office of Naval Research under grant \#N00014-20-1-2389, and the National Science Foundation under grant \#1751655. \textit{(Corresponding author: Jeffrey A. Nanzer)}}
	\thanks{	
		A. Bhattacharyya and J. A. Nanzer are with the Department of Electrical and Computer Engineering, Michigan State University, East Lansing, MI 48824 USA (email:  bhatta67@msu.edu, nanzer@msu.edu).}
}
	
	\maketitle
	
\begin{abstract}

We present a multi-node, multi-objective open-loop microwave distributed beamforming system based on high-accuracy wireless synchronization and localization. Distributed beamforming requires accurate coordination of the spatial and electrical states of the individual elements within the array to achieve and maintain coherent beamforming at intended destinations. Of the basic coordination aspects, time synchronization and localization of the elements are among the most critical to support beamforming of modulated waveforms to destinations in both the near-field and far-field of the array. In this work, we demonstrate multi-objective distributed beamforming from a three-node distributed phased array consisting of software-defined radios that leverages high-accuracy wireless time coordination for both time synchronization and two-dimensional localization of the elements. We use a spectrally-sparse two-tone waveform for high-accuracy inter-node range estimation combined with a linear-frequency modulated waveform to mitigate multipath interference. Localization is performed in a centralized format, where one node is designated as the origin and the remaining nodes build the array geometry relative to the origin, from which we obtain localization accuracy of less than \SI{1}{\centi \meter}. We implement a near-field multi-objective beamformer based on the location estimates, which enables the simultaneous steering of a beam and a null to two receiving antennas. Multi-objective beamforming of pulsed waveforms at a carrier frequency of \SI{2.1}{\giga \hertz} is demonstrated in cases where one of the nodes in the distributed antenna array is moved, and where the targets (the two receiving antennas) are moved.

\end{abstract}
	
	
\acresetall 
	
\section{Introduction}
\label{sec:intro}

Distributed wireless systems with high levels of coordination represent a paradigm shift in emerging wireless applications. Traditionally operated with only coarse cooperation between networked systems, distributed electromagnetic systems support a wide range of enhancements in wireless network capabilities: it offers a new, non-intrusive and remote method for monitoring stress, potentially providing valuable insights into individuals' well-being~\cite{1213626}, remote monitoring and management of irrigation systems, potentially leading to more efficient water usage and improved agricultural productivity~\cite{4457920}, and, it can provide a comprehensive monitoring and enhanced decision-making regarding machinery states, minimizing false alarms, and decreasing overall energy consumption in industrial settings~\cite{7576704}, among other applications. Some of the most promising capabilities lie in wireless networks with extremely accurate cooperation where nodes are coordinated at the level of the radio-frequency wavelength. Such coherent distributed systems enable distributed beamforming, affording dramatic improvements in capabilities compared to networks coordinated incoherently, and can lead to significant improvements in emerging applications such as cooperative satellite constellations for remote sensing and communications, vehicle-to-everything communications, and MIMO radar applications, among others~\cite{nasa2020taxonomy,miranda20202020,gogineni2011target}.


Coordinating individual elements in a coherent network is challenging since the relative phases, frequencies, and times of each node must be appropriately aligned at accuracies that are below that of the carrier frequency wavelength for phase and frequency, and at the level of the inverse of the information bandwidth for relative timing. Among these challenges, achieving relative phase alignment is one of the most difficult, but is also the most relevant for many applications. In remote sensing applications or communications in challenging environments, feedback from the intended destination may be minimal or absent, meaning that feedback-enabled beamforming techniques, where the relative phases can be aligned based on knowledge of the performance at or near the destination, cannot be used~\cite{quitin2013closedloop, quitin2016closedloop, peiffer2016closedloop, overdick2017closedloop}. In \cite{yan2019hybridloop}, the authors demonstrate the implementation of frequency syntonization and time synchronization for the dissemination of beamforming using bursty transmission of data packets; however, it still uses feedback from the receiver for distributed beamforming. In \cite{hanna2023hybridloop}, the authors propose a guided beamforming approach; all the nodes in the distributed array adjust their phases for coherent beamforming depending on the feedback provided by the guiding node in the array that contains relevant information about the receiver. This process is independent of the distance to the receiver, unlike in closed-loop beamforming, but still requires some prior system knowledge of the receiver. Open-loop techniques, where the array aligns itself without external inputs, allow for any wireless operation, but the relative locations of the nodes must be known to sub-wavelength accuracy to align the phases such that high levels of beamforming gain are possible~{\cite{nanzer2017openloop, mghabghab2020open}}.
In~\cite{sean2020openloop}, the use of a spectrally-sparse two-tone continuous wave signal is demonstrated for inter-node range estimation and frequency syntonization that can theoretically support coherent distributed operations up to \SI{5.96}{\giga \hertz}. In~\cite{sean2020ranging}, the authors use a pulsed two-tone stepped-frequency waveform for multinode ranging, enabling coherent beamforming up to the theoretical limit of $\sim$\SI{9.4}{\giga \hertz}.

\begin{figure}
	\centering
	\includegraphics[width=1\linewidth]{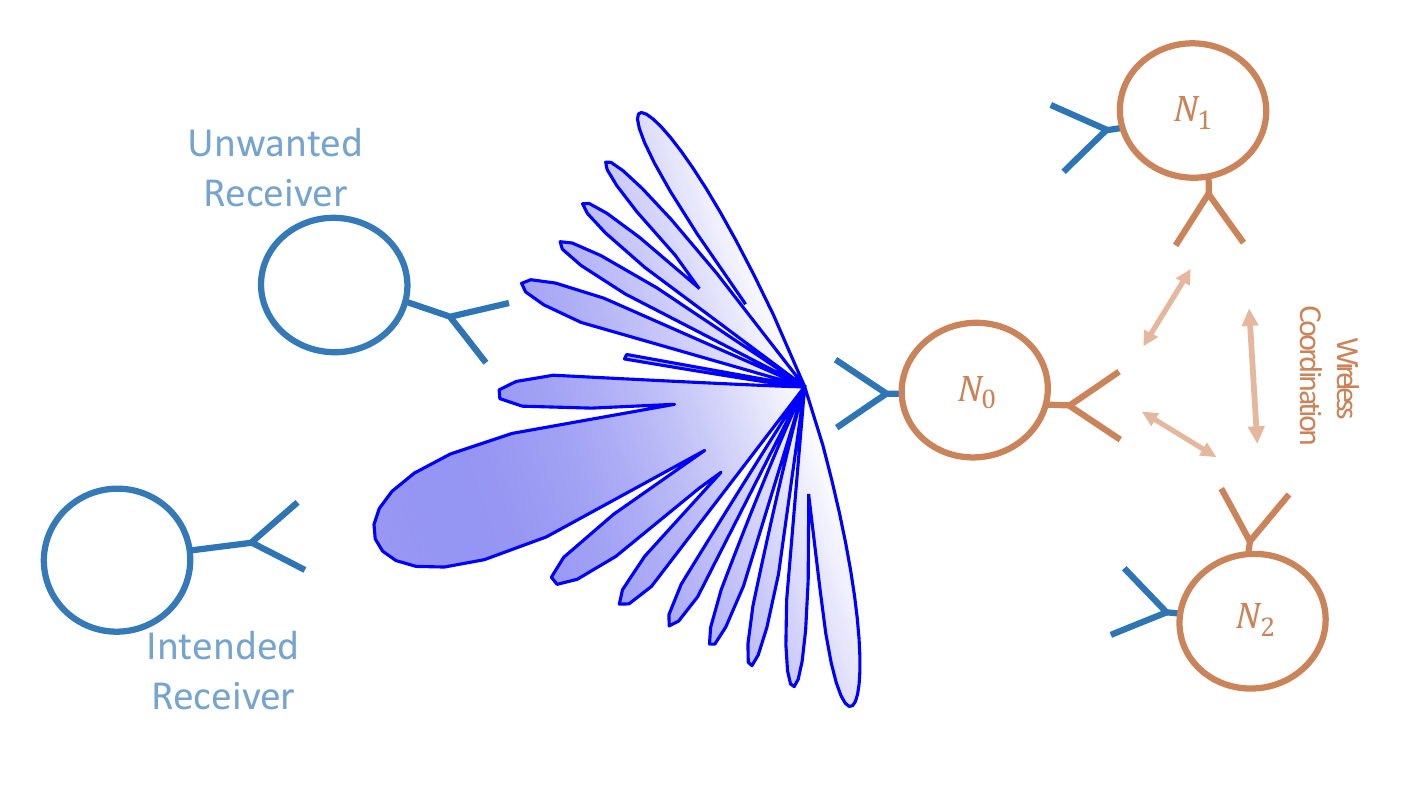}
	\caption{\textcolor{black}{Topology showing multi-objective beamforming in the near field using a three-node distributed phased array.}}
		\label{fig:daa_ov} 
	\end{figure}


In this paper we demonstrate for the first time a multi-objective distributed phased array system that achieves beamforming in the near-field of the array using high-accuracy localization and time synchronization as shown in Fig.~\ref{fig:daa_ov}. A three-element \SI{2.1}{\giga \hertz} distributed phased array based on SDRs is implemented on movable carts in an outdoor environment. The nodes leverage a recently developed two-way time synchronization approach~{\cite{merlo2022wireless}} to align the transmission times of pulsed waveforms. The time synchronization approach furthermore enables the estimation of the relative distances between the nodes. By defining one node as the origin, we geometrically localize the remaining two elements in two-dimensional space based on the relative distances between all nodes, obtaining localization accuracy below \SI{1}{\centi \meter}. We demonstrate robust performance when moving one of the nodes in the array. We experimentally demonstrate coherent combining of the transmitted pulses from each pair of nodes and from all three nodes combined. We also implement a multi-objective beamformer~{\cite{bhattacharyya2023multiobjective}} and demonstrate the ability to steer a focus and a null to two closely-spaced receiving antennas placed in the near-field of the array. We show that the focus and null can be maintained at the receiver locations are when they are moved.


\section{Localization Based on High-Accuracy Wireless Time Synchronization}
\label{sec:loc}

In this work, we estimate the relative locations of elements in a distributed network using high-accuracy estimates of the relative ranges between the elements. We designate a given node as the reference point; a second node is designated to exist along one of the axes of the two-dimensional coordinate system; from this point all subsequent nodes can ideally be localized. Estimation of the inter-node ranges is obtained by transmitting a signal between the two nodes and estimating the time of flight. In the subsection \ref{sub_sec:two-way_tt} we describe the use of a high-accuracy two-way time transfer approach for estimating inter-node ranges and simultaneously synchronizing the antenna elements in the distributed array.

\subsection{Two-Way Time Transfer}
\label{sub_sec:two-way_tt}
The synchronization of clocks on spatially distributed array nodes can be accomplished through the use of two common techniques: one-way time synchronization, where the time signal is passed from one node to the next without feedback, and two-way time synchronization where nodes exchange timing information~\cite{Levine_2008}. One-way time transfer is notably employed by global navigation satellite systems (GNSS) constellations currently in orbit. In this approach, the GNSS satellite acts as the primary clock source to which all receiver nodes synchronize by solving for signal propagation delays based on the orbital data and the known receiver position. However, to accurately determine the propagation delay, this approach requires either knowledge of both the receiver and transmitter positions, or multiple transmitting sources with known positions. In contrast, the two-way time transfer has no such constraint; it obtains accurate time synchronization without requiring knowledge of the prior locations of the the clocks, provided that all the nodes present in the system have both transmitting and receiving capabilities. It is one of the most precise methods of clock comparison and has been used for decades for satellite time transfer~\cite{hanson1989fundamentals,kirchner1991two}. 

\begin{figure}[t!]
	\centering
	\includegraphics[width=1\linewidth]{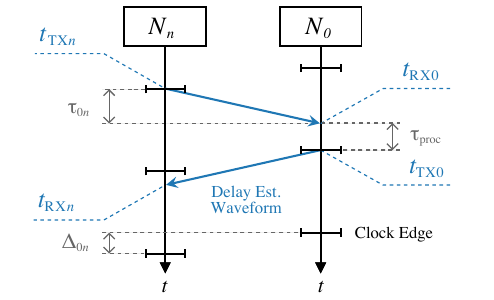}
	\caption{\textcolor{black}{Timing diagram illustrating the two-way time transfer process~\cite{merlo2022high,merlo2022wireless}. Node $N_n$  triggers the time synchronization process with the primary node, $N_0$, by transmitting a delay estimation waveform. This waveform travels from node $N_n$ to $N_0$ and back, during which timestamps are recorded at each transmission and reception. Assuming that the channels are quasi-static throughout the time synchronization epoch, these timestamps are used to evaluate the the timing bias and the propagation delay as outlined in~\eqref{eq:time-delay}}.}
	\label{fig:two_way_tt} 
\end{figure}

The heightened accuracy of the two-way time transfer is achieved through the simultaneous exchange of signals between distributed nodes. When the paths between the node clocks are nearly reciprocal or when the channel between the nodes are quasi-static in nature during the synchronization epoch, both the clock offset as well as the propagation delay can be accurately estimated to a precision on the order of \SI{1}{\micro \second}.       

The two-way time transfer approach is based on that demonstrated in~\cite{merlo2022high,merlo2022wireless} and shown in Fig.~\ref{fig:two_way_tt}. In synchronizing a distributed array, the goal is to estimate and correct the quasi-static bias or time offset which is comprised of dynamic components like frequency offset and time-varying internal delays, as well as static components such as constant system delays. The model assumes node $N_0$ as the true global time and aims to determine the bias $\Delta_{n0}$, between the primary node $N_n$ and node $N_0$. The offset between the local clocks at $N_n$ and $N_0$ is computed and then added to the local clock at $N_n$ to compensate for the the accumulated bias. Assuming the channel is quasi-static during the synchronization epoch, which is valid for epochs faster than the physical motion of the elements, the propagation delay between the nodes can be estimated by,            
\begin{equation}
	\label{eq:time-delay}
	\tau_{0n} = \frac{\left(t_{\mathrm{RX}0}-t_{\mathrm{TX}n}\right)+\left(t_{\mathrm{RX}n}-t_{\mathrm{TX}0}\right)}{2}
\end{equation}
where $t_{\mathrm{TX}i}$ and $t_{\mathrm{RX}i}$ are the times of transmission and reception at node $i$, respectively, where, $i = 0, ..., n$. Similarly, once the four timestamps are known, the relative distance between the two nodes can be estimated by
\begin{equation}
	\label{eq:time-offset}
	\delta t_{0n}=\frac{\left(t_{\mathrm{RX}0}-t_{\mathrm{TX}n}\right)-\left(t_{\mathrm{RX}n}-t_{\mathrm{TX}0}\right)}{2}
\end{equation}

As seen in \eqref{eq:time-delay} and \eqref{eq:time-offset}, the accuracy of the delay and offset estimates is dependent on the accuracy of the estimates of the four timestamps. The transmission time is generally coincident with a clock edge, which for typical SDRs has a jitter on the order of hundreds of femtoseconds, and thus the limiting factor is the ability to estimate the times of reception $t_{\mathrm{RX}i}$. To accomplish this, our prior work used a spectrally-sparse two-tone waveform that obtains near-optimal delay estimation accuracy in the order of tens of picoseconds~\cite{nanzer2017accuracy, merlo2022high, merlo2022wireless, anton2020ranging}. The benefit of the two-tone waveform is that it most closely approximates the ideal delay estimation operator, which is two discrete tones separated by a wide bandwidth. In practice the tones have some small bandwidth since they are temporally limited. Additionally, the waveform is spectrally-sparse, such that the tones can be placed at locations in the spectrum that minimize interference with other systems. Furthermore, since the bandwidth between the tones is not utilized for the estimation process, additional waveforms may be added between the tones for joint waveforms such as sensing and communications~\cite{anton2022sensing, anton2024sensing}. Previously we demonstrated the use of the two-tone waveform in the two-way time transfer algorithm described above, obtaining time synchronization accuracy of $\sim$\SI{2}{\pico \second} between two SDRs~\cite{merlo2022high, merlo2022wireless}. 
\subsection{Localization}
\label{sub_sec:loc}
One of the primary challenges in a distributed phased array is accurately estimating the locations of the elements to support phase alignment.
Previous work addressed phase alignment within open-loop systems particularly in cases where secondary nodes moved radially relative to the primary node, assuming full knowledge of their orientations (e.g.,~\cite{mghabghab2020open}). 
Ranging and localization has been extensively studied in various contexts.
In~\cite{fmcradar2018localization}, the authors demonstrate the use of sub-harmonic frequency modulated continuous wave (FMCW) radar for localization within a precision of \SI{2.97}{\centi \meter} in an indoor setting, and in~\cite{alqudsi2024fmcwlocalization} the authors discuss use of FMCW time based two-way ranging protocols for high accuracy real-time localization. In~\cite{amiri2017efficient, amiri2018eliptical}, a weighted least squares estimator is used for single-target localization in utilizing a distributed MIMO radar system. An elliptical localization approach was used in~\cite{bayat2024elliptical}, which did not require clock synchronization of the transmitters, thus trying to mitigate the complexity of the synchronization process. 
In this work, we demonstrate the capability of simultaneously aligning the times and phases of spatially separated mobile nodes of the distributed array system by leveraging the high precision two-way time transfer process.

The two-way time transfer approach in \eqref{eq:time-delay} yields the offset between the two time bases of the nodes by finding the difference between the propagation delays in the forward and return paths between. This is done by processing the received waveforms via a matched filter using a reference waveform with identical bandwidth and carrier frequency. The output of the matched filter undergoes refinement through the quadratic least squares (QLS) peak interpolation technique to mitigate discretization errors and estimate the time delay at each node. By summing the two propagation delays the resultant value is proportional to the total channel delay, from which the relative distances between the nodes can be obtained.
This time delay estimation includes $\tau_{\mathrm{cal}_{0n}}$, representing the processing time between the initial pulse reception and response at $N_n$ as well as the signal processing delays before and after digitization at the transmitter and receiver. Additionally, it includes static delays such as those through transmission lines, etc. As long as the channel is reciprocal these factors do not affect the time offset $\Delta_{n0}$. To achieve accurate range estimation, these systemic delays must be estimated and calibrated out of the resultant estimate. Because the delay $\tau_{\mathrm{cal}_{0n}}$ is dependent on system factors, it is generally deterministic, and can be estimated using an initial calibration by various means. In this paper, use a simple approach of measuring the initial true inter-node ranges to calculate $\tau_{\mathrm{cal}_{0n}}$ as explained in details in sub-section \ref{sub_sec:exp_setup}. The range can then be calculated as
\begin{equation}
	\label{eq:range}
	d_{0n} = \left( \tau_{0n} - \tau_{\mathrm{cal}_{0n}} \right)c  
\end{equation}   
where $c$ is the speed of light. 

This inter-node range can be utilized to determine the $x$ and $y$ coordinates of the array node positions. In this work we consider a three-element distributed array oriented in a two dimensional configuration and base our analysis on the following geometrical topology:
\begin{itemize} 
	\item[{a.}] The primary node is defined to be at the center of the coordinate system, that is, it is at (0, 0) coordinate. 
	\item[{b.}] The second node is defined to be located along the $x$-axis at some distance from the primary node with coordinates ($x_1$, 0).
	\item[{c.}] The third node moves freely in the first quadrant (positive $x$ and $y$ values).
\end{itemize}
The range between each pair of nodes is estimated and the Euclidean distance equation can be used to calculate the coordinates of the array nodes,
\begin{equation}
	\begin{aligned}
		d_{mn} = \sqrt{(x_m - x_n)^2 + (y_m - y_n)^2} \\
		y_n = \frac{d_{0m}^2 - d_{mn}^2 + d_{0n}^2}{2d_{0m}}\\
		x_n = \sqrt{d_{0n}^2 - y_n^2}
	\end{aligned}	
	\label{eq:loc-eq}
\end{equation}
where $(x_m, y_m)$ and $(x_n, y_n)$ are the 2-D coordinates for node $N_m$ and $N_n$ respectively. 

\section{Waveform Design and Beamforming}
\label{sub_sec:dual_lfm}

\subsection{Localization Waveform Design}
The errors in estimating the relative timing offset and relative positions of the elements are dominated by the error in estimating the time when the two signals $t_{\mathrm{RX}n}$ are received. The delay estimation waveform is based on a two-tone waveform, which obtains near-optimal delay estimation accuracy~\cite{nanzer2017accuracy}, and can be given by 
\begin{equation}
	s(t) = \alpha \, \mathrm{rect}\left(\frac{t}{T}\right)\left[ e^{j 2\pi \left(f - \frac{\Delta f}{2}\right) t} + e^{j 2 \pi \left(f + \frac{\Delta f}{2}\right) t} \right]  
	\label{eq:tt_wfm}
\end{equation}
where $\alpha$ is the amplitude, $\mathrm{rect}\left(\cdot\right)$ is the rectangular pulse function, $T$ is the pulse width, $\Delta f$ is the tone separation, $t$ is the total duration and $\phi$ is the phase of the waveform respectively.
Generally, the two-tone waveform exhibits appreciable ambiguities in range, due to the periodicity of the temporal shape of the waveform. Such ambiguities cause the two-tone waveform to be relatively intolerant to low SNR ot strong fading multipath effects, which is problematic for distributed phased arrays since even a minor bias in the time delay estimation on the order of nanoseconds can result in position bias errors on the order of tens of centimeters, severely limiting the effectiveness of a distributed beamforming operation. 

To address these challenges, we combined the two-tone waveform with the linear frequency modulation (LFM) waveform of bandwidth $\Delta f_{\mathrm{LFM}}$ to create a hybrid waveform referred to as the dual linear frequency modulated (dual-LFM) waveform~\cite{9329928,bhattacharyya2023dual} which is given by
\begin{equation}
s(t) = \alpha\,\mathrm{rect}\left(\frac{t}{T}\right)e^{j\frac{\pi}{T}\Delta f_{\mathrm{LFM}}t^2}e^{j2\pi f t}
\label{eq:dua_lfm_wfm}
\end{equation}
where $\tau$ and $t$ are the pulse duration of the LFM waveform and the total waveform respectively. In this paper we consider $\Delta f_{\mathrm{LFM}}$ to be equal to half of the total bandwidth $\Delta f$.
Like the traditional LFM, the dual-LFM is unambiguous, however at any instantaneous time the waveform is represented at a two-tone signal. 

\begin{figure}[t!]
	\centering
	\includegraphics[width=1\linewidth]{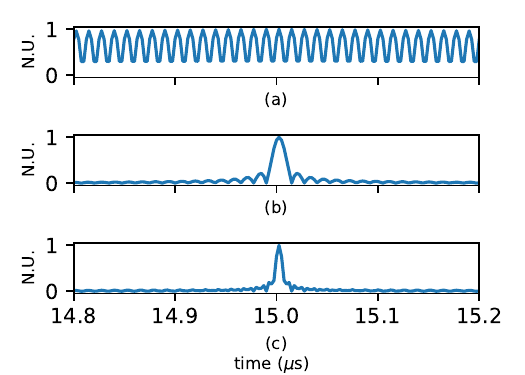}
	\caption{Example of matched filter output for (a) two-tone, (b) LFM, and (c)
		dual-tone LFM waveform showing the changes in the side-lobe level
		between the three waveforms. (N.U.: Normalized units)}
	\label{fig:mf_wfms} 
\end{figure}

The received waveform undergoes matched filtering with a known zero-delay reference, which maximizes the SNR at the output of the filter. The two-tone waveform is ambiguous in time, as seen in the matched filter response in Fig.~\ref{fig:mf_wfms}(a), where the high time sidelobes are clear. In contrast, the matched filter response of the linear frequency modulation (LFM) waveform, illustrated in Fig.~\ref{fig:mf_wfms}(b), has significantly lower sidelobe levels, but also has a broader peak, yielding unambiguous but less accurate time estimates than the two-tone waveform.
The matched filter output of the dual-LFM is shown in Fig.~\ref{fig:mf_wfms}(c), where it is clear that lower sidelobe levels are obtained compared to the two-tone waveform, but that the peak is narrower. Previously, we demonstrated that this waveform could obtain localization accuracy of \SI{4.2}{\milli \meter} and synchronization accuracy of \SI{4.7}{\pico \second}, commensurate with supporting coherent beamforming at frequencies up to \SI{4.4}{\giga \hertz}~\cite{bhattacharyya2023dual}.

\begin{figure*}[t!]
	\centering
	\includegraphics[width=0.8\linewidth]{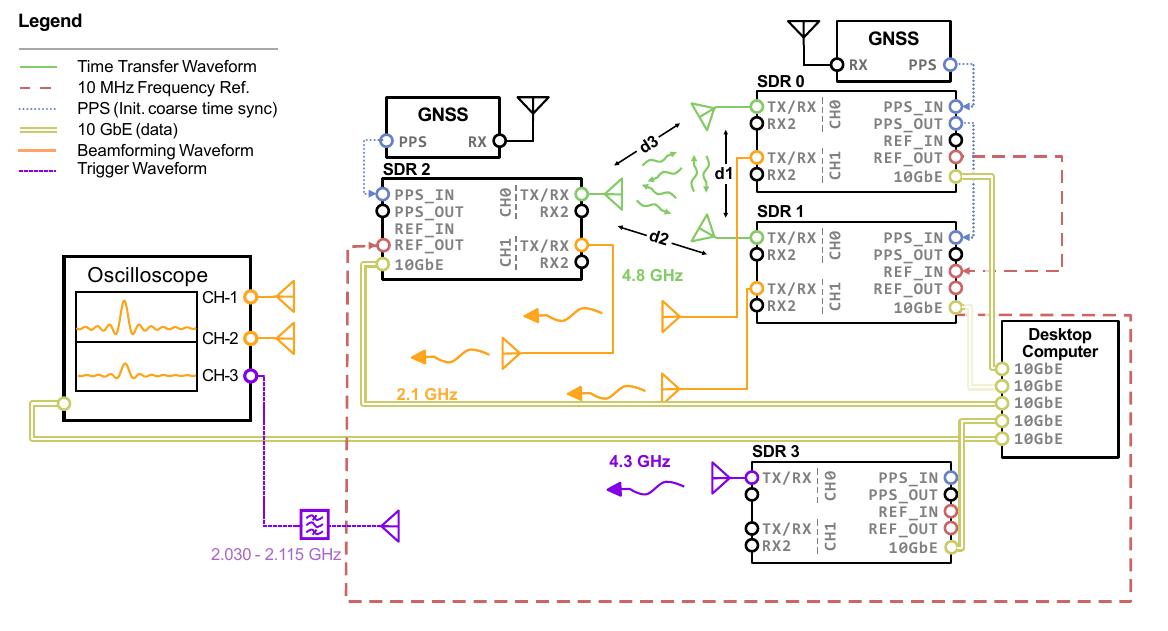}
	\caption{Wireless time transfer, cabled frequency transfer near-field multi-objective beamforming system schematic. The experiment configuration shows SDRs forming the three nodes of the distributed array. SDR 0 acts as the primary SDR and the frequency syntonization is done using the internal \SI{10}{\mega \hertz} reference from SDR 0, distributed using coaxial cables. A pulse-per-second (PPS) is used to initialize the time synchronization epoch window using GNSS antenna on SDR 0 and SDR 2. The same is done via a coaxial cable connecting SDR 0 and SDR 1. The oscilloscope is utilized to sample and digitize the beamforming waveforms, enabling the determination of beamforming accuracy during the execution of wireless localization. SDR 3 is used as an auxillary node to trigger the oscilloscope. A narrow-band filter was employed on the triggering receiver to eliminate undesired RF signals, ensuring that the oscilloscope would exclusively capture pulses transmitted from the array. This operation remained independent of the beamforming performance. The 10 Gb Ethernet cables are employed to control the nodes and the oscilloscope through GNU Radio software.}
	\label{fig:schematic} 
\end{figure*}



\subsection{Near-Field Multi-Objective Beamforming Algorithm}
\label{sec:lcmp_nf_bf}
Prior work in multi-objective beamforming mostly include closed-loop systems~\cite{kumar2015nulforming, goguri2016mobeamforming, peiffer2017nullforming} and focused on far-field beam and null steering.
In~\cite{bhattacharyya2023multiobjective}, distributed phased array multi-objective beamforming in the near field was explored based on wireless frequency syntonization, but without wireless localization or time synchronization. This prior work also used only continuous-wave signals. Here we demonstrate the complementary facets, using high-accuracy localization and synchronization for multi-objective distributed beamforming in the near field using pulsed waveforms. 
The multi-objective beamforming algorithm adapts the traditional quiescent pattern linear constrained minimum power (LCMP) far-field multi-objective beamforming algorithm detailed in~\cite[p.~513--553]{trees2002optimum} for beamstearing in the near field. 

The conventional LCMP method involves using a null constraint matrix $\mathbf{C}$ to compute the necessary weights for multi-objective beamforming. The matrix $\mathbf{C}$ is an $N \times M$ matrix where $N$ represents the number of transmitters and $M$ signifies number of receiving locations towards which beams or nulls are directed. Hence, the columns of the matrix $\mathbf{C}$ are linearly independent from each other. The weights for beamforming and null forming are determined by
\begin{equation}
	\label{eq:w_lcmp}
	\mathbf{w}_\mathrm{lcmp} = \mathbf{g}^H(\textbf{C}^H\textbf{C})^{-1}\textbf{C}^H
\end{equation} 
where  $(\cdot)^H$ represents the Hermitian operation. The constraint matrix $\mathbf{C}_{N \times M}$ consists of vectors containing the channel weights of each receiver, which are determined by \mbox{$\mathbf{v}_{m,n} = e^{-jkr_{mn}}$}, where $r_{mn}$ represents the distance between transmitters and receiver locations given by,
\begin{equation}
	\label{eq:range_lcmp}
	r_{mn} = \sqrt{\left( x_\mathrm{TX_n} - x_\mathrm{RX_m} \right)^{2} + \left( y_\mathrm{TX_n} - y_\mathrm{RX_m} \right)^{2}} .
\end{equation}
The localization approach is used to estimate the 2D coordinates of the transmit nodes $x_\mathrm{TX_n}$ and $y_\mathrm{TX_n}$. The beam or the null can be steered to any given ($x_\mathrm{RX_m}, y_\mathrm{RX_m}$).
The vector g specifies beams $B_m=1$ or nulls $B_m=0$ at the receiver locations. The relative positions of the receiving elements are arbitrarily set and the transmitter locations are determined using \eqref{eq:loc-eq} through high-accuracy range estimation provided by the the two-way time synchronization process using dual-LFM waveform. 

\section{Near-Field Beamforming Experimental}
\label{sec:nf_setup}
\subsection{Experimental Configuration}
\label{sub_sec:exp_setup}
\begin{figure*}
	\centering
	\includegraphics[width=1\linewidth]{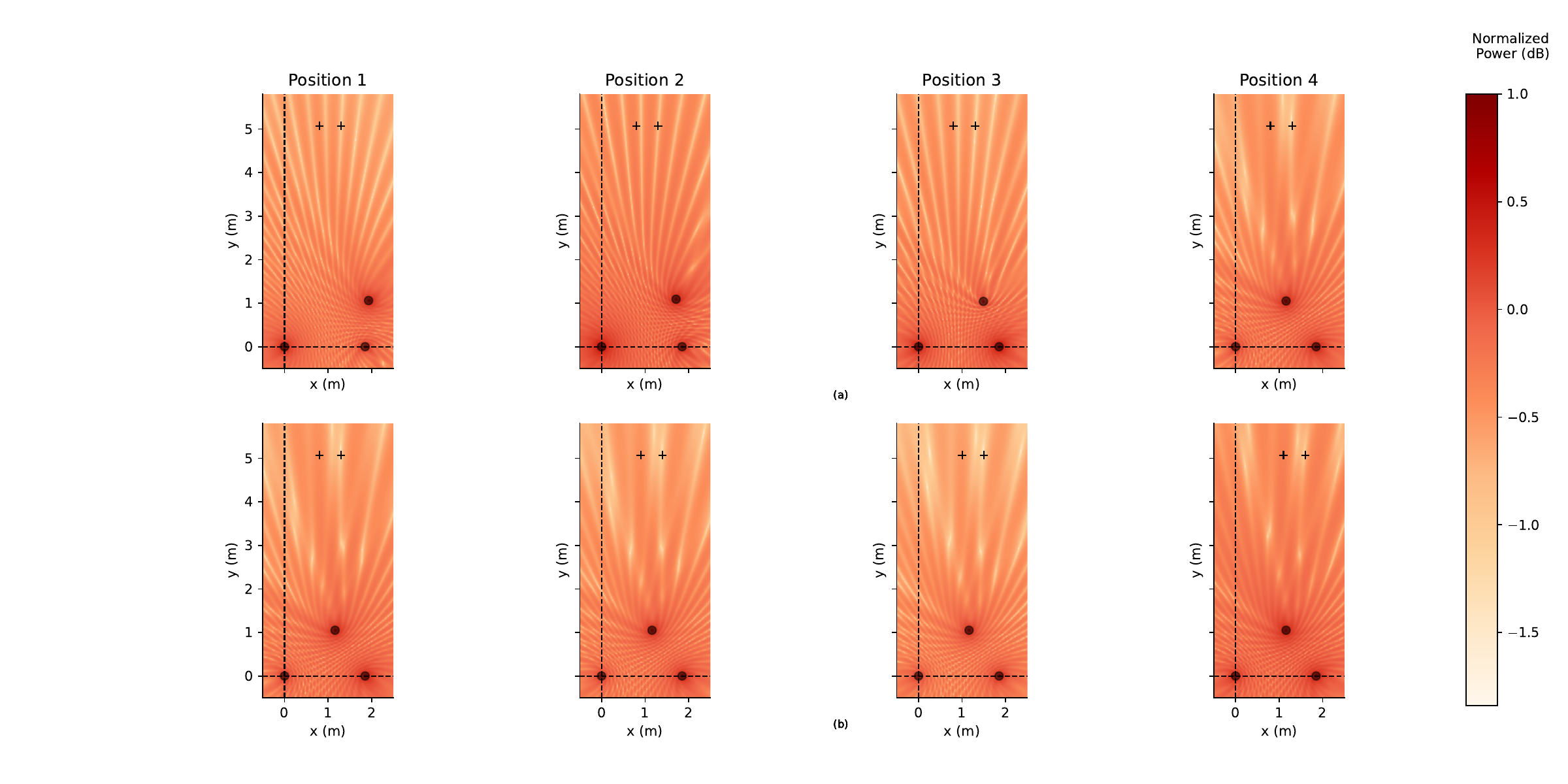}
	\caption{Simulation of the near-field beamforming experiment (a) with spatially dynamic transmitter (node 2) and (b) beamsteering to a moving target locations with static transmitters. The left target locations is a focus while the right location is a null.}
	\label{fig:exp_setup_sima} 
\end{figure*}

In the near-field multi-objective beamforming experiments, three SDRs, SDR 0, SDR 1 and SDR 2 represents the three separate nodes of the distributed array. The two experiments conducted were:
\begin{itemize}
	\item[{a.}] dynamic transmit node: SDR 2 is moved arbitrarily to five different positions including the calibration position. The receivers are kept at the same locations. Beamforming is not performed at the calibration location of the nodes. 
	\item[{b.}] beam and null steering: The transmitting nodes are static while the receivers are moved to four different locations to show beam and null steering. 
\end{itemize}
Fig.~\ref{fig:exp_setup_sima} shows the results of the simulations of the experiments. The dashed black lines indicate the $x = 0$ and  $y = 0$ axes of the 2D coordinate system. The primary Node 0 (SDR 0) is placed at $(x = 0, y = 0)$. Node 1 (SDR 1) is defined to lie at a distance of d1 from Node 0 along the $x$-axis. Node 2 (SDR 2) is placed and moved arbitrarily. The cross signs point to the locations of receiving antennas. The beamforming simulations steer a focus to the left receiver and a null to the right receiver. The top row shows the beamforming performance as Node 2 is moved, showing that the focus and null maintained at the receiver for a spatially dynamic node. The bottom row shows the beamforming performance when the two receivers are moved, again showing that the focus and null maintained.

\begin{figure}
	\centering
	\includegraphics[width=1\linewidth]{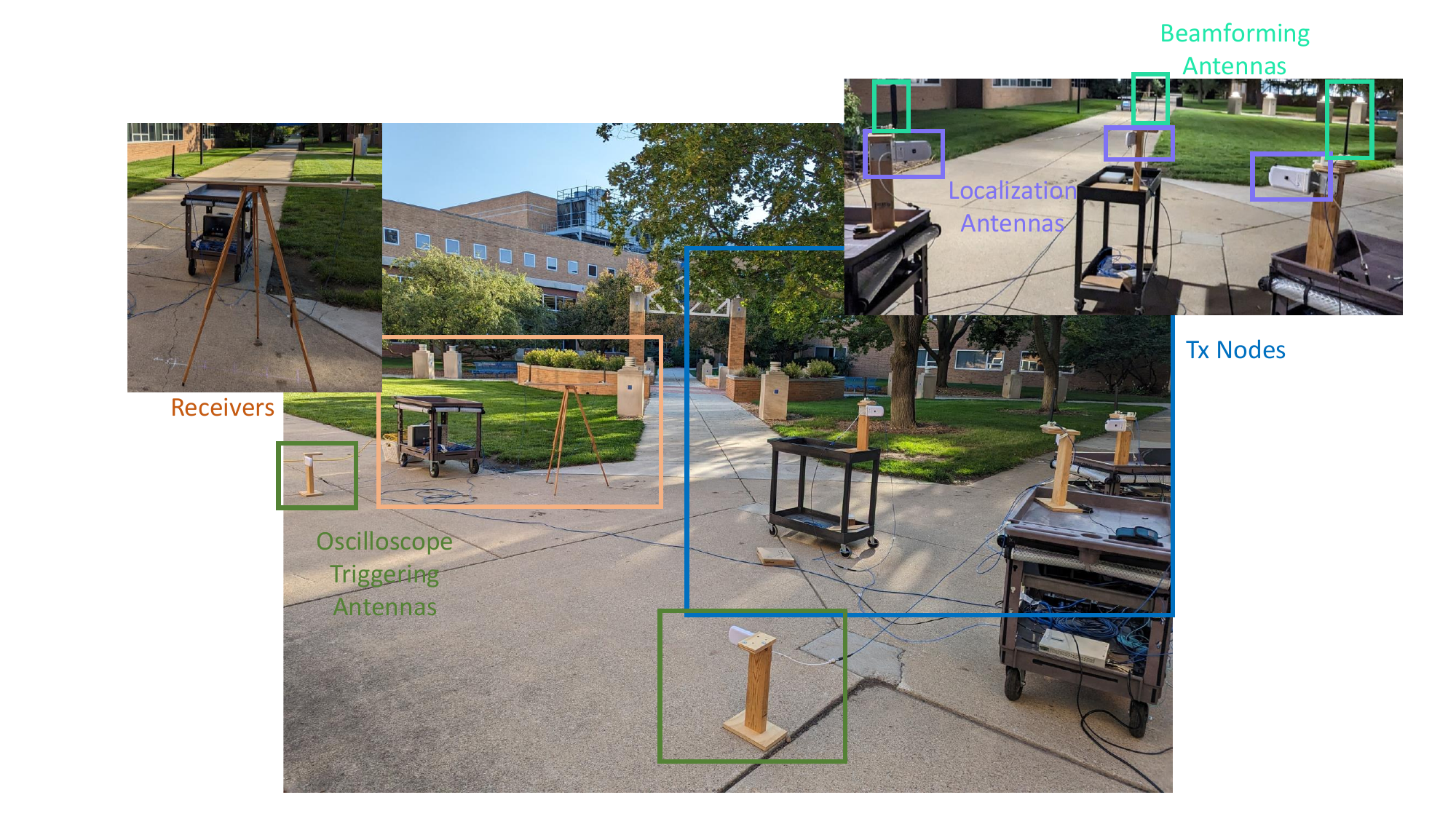}
	\caption{Photographs of the experimental outdoor setup.}
	\label{fig:exp_setp} 
\end{figure}

A block diagram of the experimental setup is shown in Fig.~\ref{fig:schematic}, and an photograph of the setup in an outdoor testing environment is shown in Fig.~\ref{fig:exp_setp}. The setup consisted of three Ettus Research USRP X310 SDRs each equipped with two UBX-160 daughterboards. Each of the SDRs were equipped with an L-Com Infinite brand Ultra-Wideband \SI{8}{~dBi} log periodic antenna (\SI{2.4}{} to \SI{6.5}{\giga \hertz}) for wireless localization, and a PulseLarsen SPDA24700/2700 dipole antenna for near-field multi-objective beamforming. Antennas with directionality were chosen to help minimize potential ground bounce multipath. The beamformed signals were received on two antennas connected to two channels of an oscilloscope. As the main objective of this work was to evaluate the performance of localization and synchronization in particular, the nodes were syntonized via cable. Note that wireless syntonization may be accomplished in various ways~{\cite{mghabghab2020self, mghabghab2020open, merlo2022wireless}}. The SDRs were controlled using GNU Radio software on a PC connected to each node via 10 Gb Ethernet; this connection may also be implemented directly using various wireless communications protocols. Initial coarse alignment of the time bases on the nodes was achieved within 100 ns through GNSS pulse-per-second synchronization. The dual-LFM waveform with a pulse duration of \SI{10}{\micro \second} at \SI{4.8}{\giga \hertz} and instantaneous bandwidth of \SI{40}{\mega \hertz} was transmitted and received by the log periodic antennas between each of the nodes. The sampling rate on the receivers of the SDRs was \SI{200}{\mega Sa/ \second}. The dual-LFM waveform was used for both synchronization and localization via \eqref{eq:time-delay} and \eqref{eq:time-offset}, respectively. Ground truth measurements were obtained using a laser rangefinder. The nodes were arranged in an arbitrary triangular orientation where the distance between SDR 0 and SDR 1 is d1, the distance between SDR 1 and SDR 2 denoted as d2 and the distance between SDR 0 and SDR 2 is d3. The initial position is the calibration position that is used to determine $\tau_{\mathrm{cal}_{0n}}$ for~\eqref{eq:range}.

\begin{figure}
	\centering
	\includegraphics[width=1\linewidth]{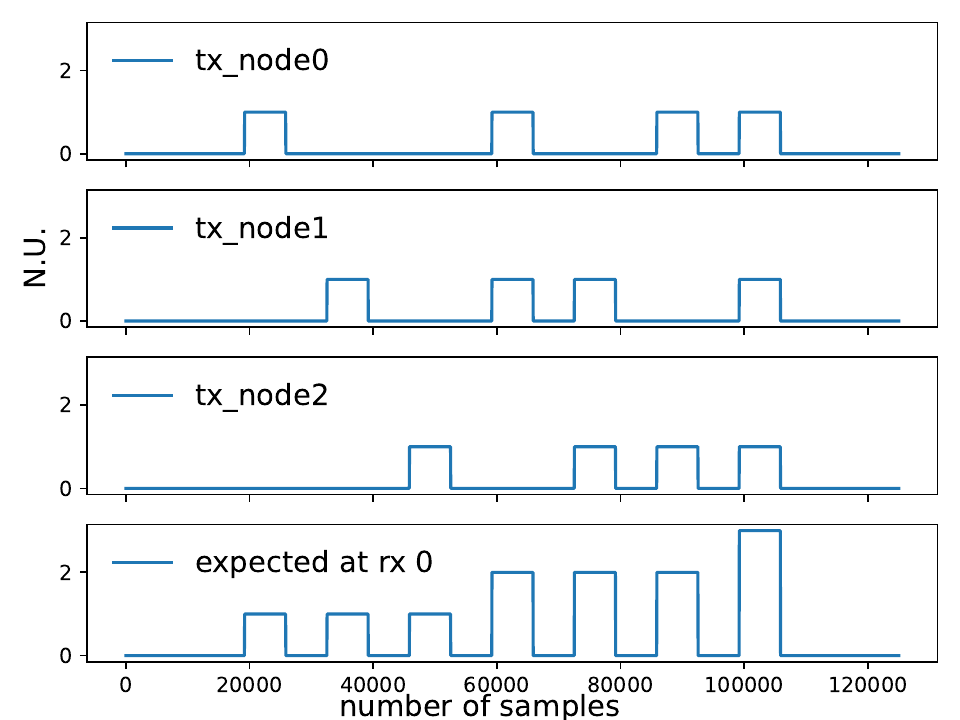}
	\caption{Simulation of the envelopes of the transmitted waveforms. Each node transmits a sequence of pulses in a pattern that yields reception at the focal location of the three transmitters individually, followed by each pair of transmitters, followed by the summation of all three. This pattern allows a quantitative evaluation of the relative beamforming gain.}
	\label{fig:tx_ask_bf_sig} 
\end{figure}

Channel 0 on each of the SDRs is used for real time localization while channel 1 is used for beamforming. To calibrate the static time delays introduced by internal hardware paths and processing latency, channel 1 of the SDRs was directly connected to the oscilloscope via a coaxial cable. Each node transmitted a \SI{40}{\mega \hertz} bandwidth LFM signal at a carrier frequency of \SI{2.1}{\giga \hertz}. The phase calibration process then follows similarly to the propagation delay estimation process detailed in \ref{sub_sec:loc}. The waveforms received by the oscilloscope were matched filtered and then refined using a QLS to estimate the time and phase delay at each node. This data processing was done offline at the beginning of the experiment. 
Utilizing this information, the delay of the primary nodes were subtracted from the other two nodes, calibrating the entire system with respect to the primary node for subsequent beamforming. The 2D coordinates obtained from the localization process, \eqref{eq:loc-eq}, was used to obtain range between the transmitters and the receivers using \eqref{eq:range_lcmp}. In these experiments the distance between the time synchronization antennas and the beamforming antennas are known, thus the difference is manually corrected directly in the GNU radio. These calculated ranges were then used to obtain the LCMP weights \eqref{eq:w_lcmp}. These weights are finally used to beamform at channel 1 of the oscilloscope and nullform at channel 2. 

The beamforming signals were transmitted by each SDR on dipole antennas. Each node transmitted an amplitude shift key (ASK) signal at \SI{2.1}{\giga \hertz} carrier frequency with a bandwidth of \SI{40}{\mega \hertz} and pulse duration of \SI{10}{\micro s} at a rate of \SI{1.5}{\mega Sa/ \second}. The pulse pattern for each transmitter was staggered to allow for the reception of waveforms from each individual node as well as all combinations of the transmitted waveforms between each node pairwise and for all three combined, as shown in Fig.~\ref{fig:tx_ask_bf_sig}. Parameters of the ASK waveforms are given in Table \ref{tab:wfm_param}. The $*$ denotes complex signal sampling rate. This combinaton of signals allowed an evaluation of the distributed beamforming performance by directly comparing the combined and individual transmitter responses. The parameters for both the waveforms and antennas are given in \ref{tab:wfm_param}.

\begin{table}[tb]
	\ifcsname hlon\endcsname%
	\color{blue}%
	\fi%
	\caption{{Experiment Waveform Parameters}}
	\label{tab:wfm_param}
	\begin{center}
		\begin{tabularx}{\columnwidth}{p{0.4\linewidth}Y}
			
			\toprule[1pt]
			\textbf{Time Transfer Waveform} & \\
			\midrule
			\midrule
			Parameter & Value \\
			\midrule
			Waveform Type & \mbox{Dual\;LFM} \\
			Carrier Frequency & \SI{4.8}{\giga \hertz} \\
			Bandwidth & \SI{40}{\mega \hertz} \\
			Pulse Duration & \SI{10.0}{\micro \second} \\
			Rise and Fall Time & \SI{5}{\nano \second} \\
			Rx Sample Rate & \SI{200}{\mega Sa/\second}* \\
			Tx Sample Rate & \SI{200}{\mega Sa/\second}* \\
			\midrule[1pt]
			\textbf{Beamforming Waveform} & \\
			\midrule
			\midrule
			Parameter & Value \\
			\midrule
			Waveform Type & \mbox{ASK} \\
			Carrier Frequency & \SI{2.1}{\giga\hertz} \\
			Bandwidth & \SI{40}{\mega\hertz} \\
			Pulse Duration & \SI{10.0}{\micro\second} \\
			Rise and Fall Time & \SI{5}{\nano \second} \\
			Tx Sample Rate & \SI{200}{\mega Sa/\second}* \\ 
			Tx 0 data & [0,1,0,0,0,0,0,1,0,0,0,1,0,1,0] \\
			Tx 1 data & [0,0,0,1,0,0,0,1,0,1,0,0,0,1,0] \\
			Tx 2 data & [0,0,0,0,0,1,0,0,0,1,0,1,0,1,0] \\
			Data Rate & \SI{1.5}{\mega Sa/\second} \\
			Rx Sample Rate & \SI{10}{\giga Sa/\second} \\
			\midrule[1pt]
			\textbf{Oscilloscope Triggering Waveform} & \\
			\midrule
			\midrule
			Parameter & Value \\
			\midrule
			Waveform Type & \mbox{CW} \\
			Carrier Frequency & \SI{4.3}{\giga\hertz} \\
			Bandwidth & \SI{50}{\mega\hertz} \\
			Pulse Duration & \SI{10.0}{\micro\second} \\
			Rise and Fall Time & \SI{10}{\nano \second} \\
			Tx Sample Rate & \SI{200}{\mega Sa/\second}* \\ 
			Rx Sample Rate & \SI{10}{\giga Sa/\second} \\
			\midrule[1pt]
			\textbf{Log-Periodic Antenna Parameters} &\\
			\midrule
			\midrule
			Parameter & Value \\
			\midrule
			Gain & $8$\,dBi \\
			Bandwidth &  \SI{2.3}{}--\SI{6.5}{\giga\hertz}\\
			\midrule[1pt]
			\textbf{Dipole Antenna Parameters} &\\
			\midrule
			\midrule
			Parameter & Value \\
			\midrule
			Gain & $2.8$\,dBi \\
			Bandwidth &  \SI{2.3}{}--\SI{6.5}{\giga\hertz}\\			
			\bottomrule[1pt]	
		\end{tabularx}
	\end{center}
\end{table}

\subsection{Experimental Results}
\label{sec:results}

Experiments were conducted to demonstrate simultaneous focusing and nulling at the two receiving antennas. The phase and amplitude of the signals emitted by the three nodes are determined by the LCMP algorithm. The focus was directed to an antenna indicated as RX 0 while the null was directed to RX 1. 
In the two experiments, either Node 2 was moved while maintaining the focus and null, or the receiver antennas were moved.

In experiment (a), the mobile node (Node 2) of the distributed array system is moved to four different locations excluding the calibration location. The positions of the nodes and receivers in each experiment are given in Table \ref{tab:node_pos}.
\begin{table*}[htbp]
	\ifcsname hlon\endcsname%
	\color{blue}%
	\fi%
	\caption{{Experiment Setup}}
	\label{tab:node_pos}
	\begin{center}
		\begin{tabularx}{\linewidth}{p{0.3\linewidth}Y Y Y Y Y}
			\toprule[1pt]
			\textbf{Experiment A: Moving Node 2} & & & & \\
			\midrule
			\midrule
			Transceiver & Calibration & Position 1 & Position 2 & Position 3 & Position 4 \\
			\midrule
			Node 0 & (0, 0)  & (0, 0)& (0, 0) & (0, 0) & (0, 0) \\
			Node 1 &  (0, \SI{1.85}{m}) & (0, \SI{1.85}{m}) & (0, \SI{1.85}{m}) & (0, \SI{1.85}{m}) & (0, \SI{1.85}{m})\\
			Node 2 &  (\SI{1.53}{m}, \SI{1.03}{m}) & (\SI{1.93}{m}, \SI{1.06}{m}) & (\SI{1.71}{m}, \SI{1.09}{m}) & (\SI{1.49}{m}, \SI{1.04}{m}) & (\SI{1.16}{m}, \SI{1.05}{m})\\
			RX 0 &  (\SI{0.8}{m}, \SI{5.07}{m}) & (\SI{0.8}{m}, \SI{5.07}{m}) & (\SI{0.8}{m}, \SI{5.07}{m}) & (\SI{0.8}{m}, \SI{5.07}{m}) & (\SI{0.8}{m}, \SI{5.07}{m})\\
			RX 1 & (\SI{1.3}{m}, \SI{5.07}{m})  & (\SI{1.3}{m}, \SI{5.07}{m}) & (\SI{1.3}{m}, \SI{5.07}{m}) & (\SI{1.3}{m}, \SI{5.07}{m}) & (\SI{1.3}{m}, \SI{5.07}{m})\\
			\midrule[1pt]
			\textbf{Experiment B: Beam and Null Steering} & & & &\\
			\midrule
			\midrule
			Transceiver &  & Position 1 & Position 2 & Position 3 & Position 4 \\
			\midrule
			Node 0 &  & (0, 0)& (0, 0) & (0, 0) & (0, 0)\\
			Node 1 &  & (0, \SI{1.85}{m}) & (0, \SI{1.85}{m}) & (0, \SI{1.85}{m}) & (0, \SI{1.85}{m})\\
			Node 2 &  & (\SI{1.16}{m}, \SI{1.05}{m}) & (\SI{1.16}{m}, \SI{1.05}{m}) & (\SI{1.16}{m}, \SI{1.05}{m}) & (\SI{1.16}{m}, \SI{1.05}{m})\\
			RX 0 &   & (\SI{0.8}{m}, \SI{5.07}{m}) & (\SI{0.9}{m}, \SI{5.07}{m}) & (\SI{1.0}{m}, \SI{5.07}{m}) & (\SI{0.1}{m}, \SI{5.07}{m})\\
			RX 1 &  & (\SI{1.3}{m}, \SI{5.07}{m}) & (\SI{1.4}{m}, \SI{5.07}{m}) & (\SI{1.5}{m}, \SI{5.07}{m}) & (\SI{1.6}{m}, \SI{5.07}{m})\\
			\bottomrule[1pt]	
		\end{tabularx}
	\end{center}
\end{table*}

\begin{figure}
	\centering
	\includegraphics[width=1\linewidth]{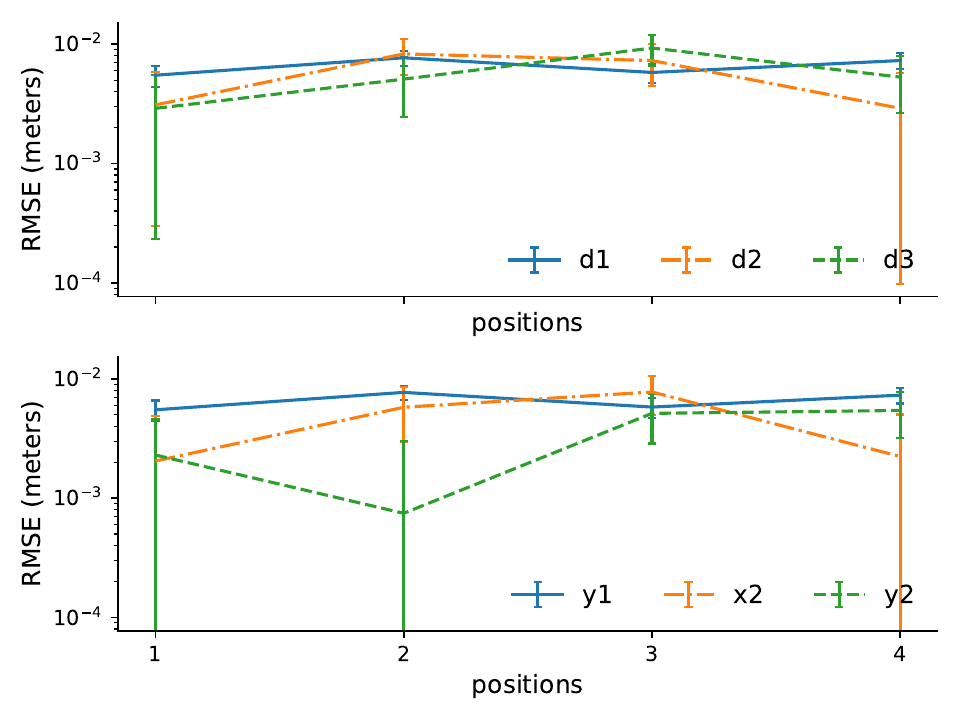}
	\caption{Root mean square error (RMSE) and standard deviation (error bars) of the estimated range (top) and estimated unknown locations (bottom) for experiment (a) with moving node 2.}
	\label{fig:tx_dy_loc_err} 
\end{figure}

\begin{figure}
	\centering
	\includegraphics[width=1\linewidth]{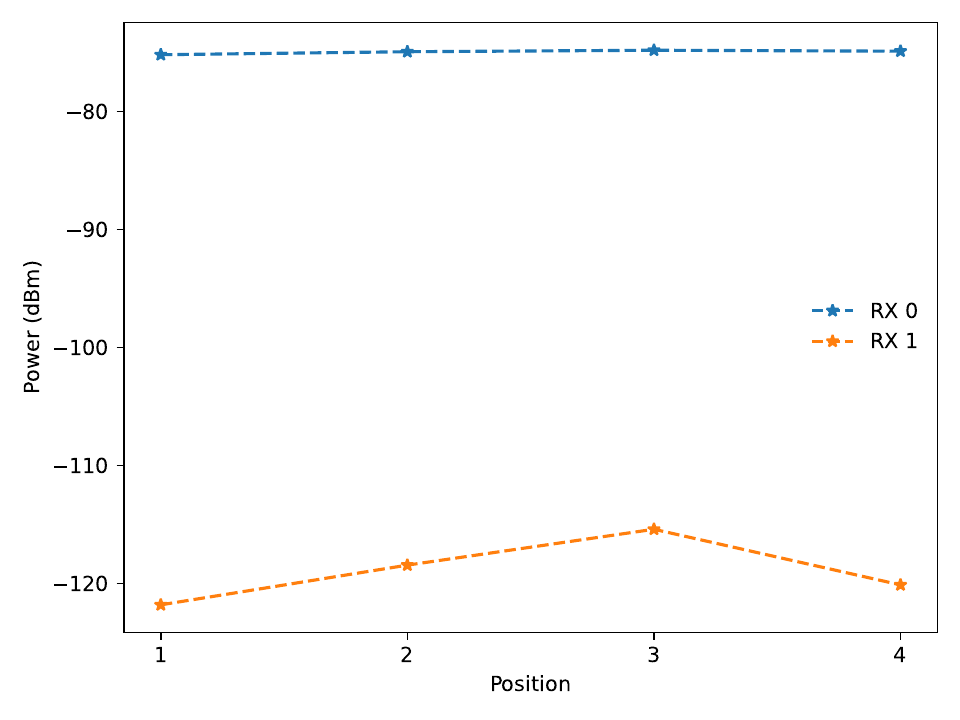}
	\caption{Average received power at the focus RX 0 and null RX 1 while Node 2 was moved to the four different positions, demonstrating the ability to maintain simultaneous focusing and nulling.}
	\label{fig:avg_rx_pw_dtx} 
\end{figure}

\begin{figure}
	\centering
	\includegraphics[width=1\linewidth]{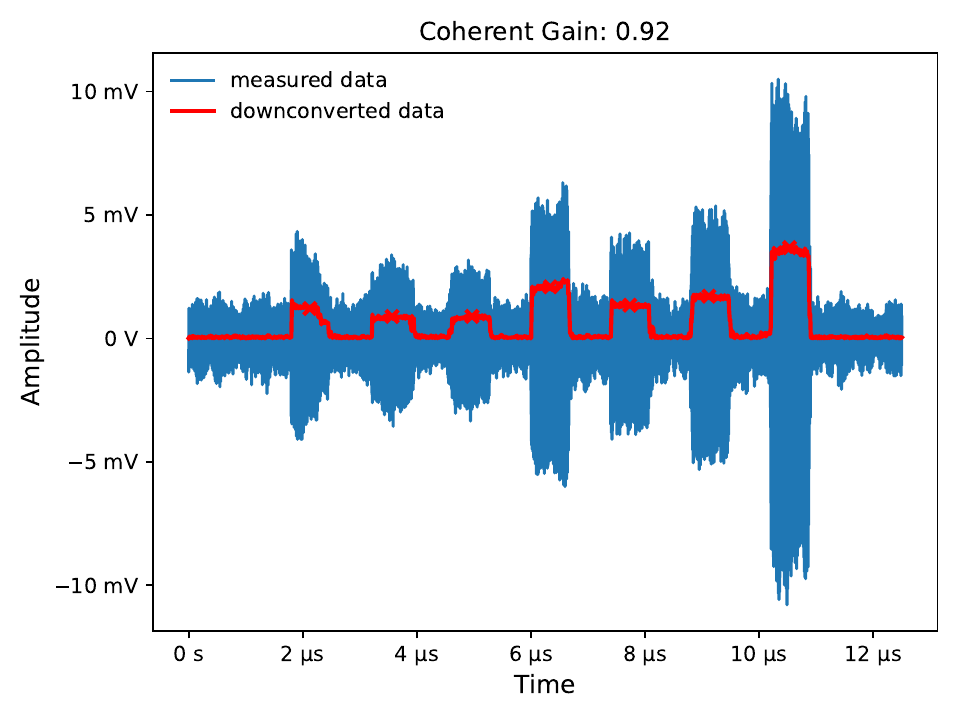}
	\caption{Example of average received waveform at the receiver RX 0. The coherent gain in this example was \SI{92}{\percent}.} 
	\label{fig:rx_bf_sig} 
\end{figure}

The error in estimating the internode distance and the error in the resultant localization of Node 2, along with error bars representing the standard deviation, are shown in Fig.~\ref{fig:tx_dy_loc_err}. 
The RMSE is calculated between the range estimated using two-way time transfer and the ground truth measured by the laser range finder and is found to be \SI{4.7}{\milli \meter} on an average for d1, d2 and d3, for the four different location configurations. The accuracy is $\sim$ \SI{2}{\milli \meter} given by the standard deviation. The average bias of the range estimate is \SI{5.8}{\milli \meter}. As stated in section \ref{sub_sec:loc} the 2-D coordinates of the mobile node 2 can be calculated using~\eqref{eq:loc-eq} and it is assumed that node 0 and node 1 act as anchors. The 2-D coordinates calculated from the range measured by the laser range finder are considered to be the reference for calculating the RMSE for $y_1$, $x_2$ and $y_2$.
In Fig.~\ref{fig:avg_rx_pw_dtx} the average power at RX 0 and RX 1 are plotted for all the four different locations of Node 2, showing clearly that the distributed beamforming system maintains a focus. The power level at RX 1 is lower by almost \SI{33.3}{\percent} than RX 0. This shows that the beamforming system can support multi-objective beamforming with moving nodes. Fig.~\ref{fig:rx_bf_sig} shows an example of the received signals at RX 0, illustrating a clear match to the expected signal shape indicated in the simulations of Fig.~\ref{fig:tx_ask_bf_sig}. 



\begin{figure}
	\centering
	\includegraphics[width=1\linewidth]{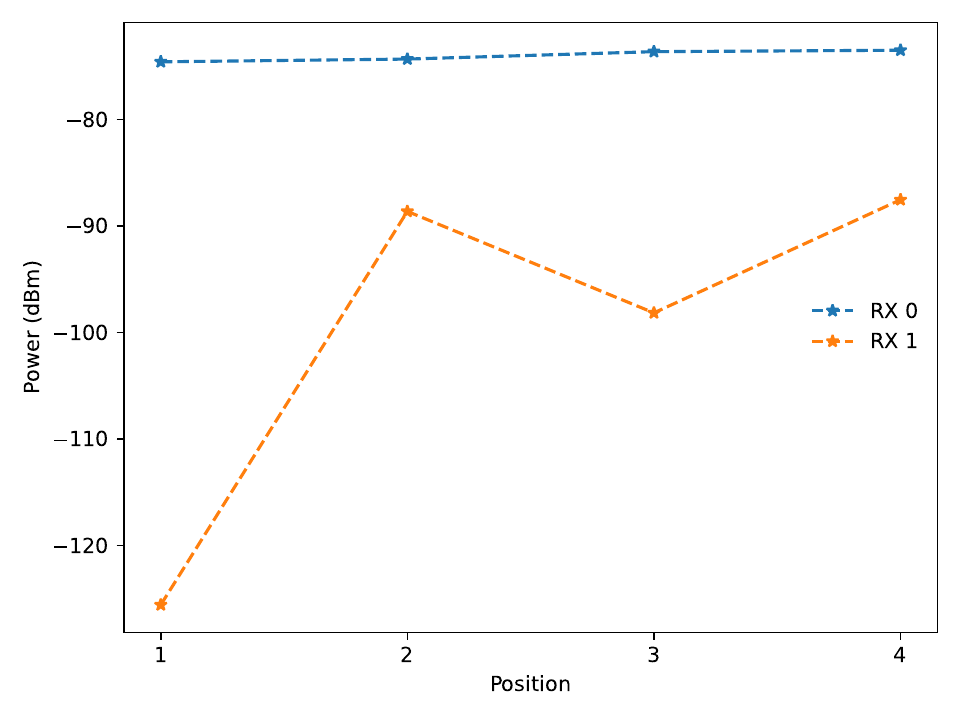}
	\caption{Average received power at the focus RX0 and null Rx1 demonstrating beamsteering in experiment (b) with moving target locations.}
	\label{fig:avg_rx_pw_bs} 
\end{figure}

Fig.~\ref{fig:avg_rx_pw_bs} shows the multi-objective beamforming results as the receiver is moved to the four different positions, indicated as experiment (b). The power at the focus remains strong and constant, and commensurate with experiment (a). The depth of the null is reduced compared to experiment (a), due in large part to multipath. However, the null depth is still consistently greater than 15 dB, indicating that the beamforming system can support multi-objective beamforming and beamsteering towards different locations.


\section{Conclusion}
In this paper we demonstrated the feasibility of multi-objective distributed beamforming with beamsteering using a 3-node distributed phased array with wireless synchronization and localization. The LCMP algorithm was used to compute weights for simultaneous focusing and nulling, based on the estimated locations of the elements in two dimensions. Experiments demonstrate that the high-accuracy synchronization and localization approach provides sufficient accuracy to maintain high levels of coherent gain in the presence of moving nodes or when beam and null steering to different locations. Future work will combine these results with wireless frequency syntonization for a fully wireless multi-objective distributed beamforming system.

\bibliography{biblatex_all.bib}
\bibliographystyle{IEEEtran}

\end{document}